# Pressure induced magnetic-field-free superconducting diode effect in $NbSe_2$ flake

Shihao Zhu,[1,†] Tian Le,[2,3,†] Cuiying Pei,[1,†] Changhua Li,[1,†] Yi Liao,[1] Yi Zhao,[1] Lingxiao Zhao,[1] Qi Wang,[1,4] Juefei Wu,[1] Qilian Zhang,[5] Yueshen Wu,[1] Tonghuan Fu,[6] Xujie Lü,[6] Wenge Yang,[6] Jie Shen,[7] Jun Li,[1,4] Yulin Chen,[1,4,8] Xiao Lin,[9,*] Wen-Yu He,[1,*] Yanpeng Qi[1,4,10,*]

[1]State Key Laboratory of Quantum Functional Materials, School of Physical Science and Technology, ShanghaiTech University, Shanghai 201210, China
[2]Center for Quantum Matter, School of Physics, Zhejiang University, Hangzhou 310058, China
[3]Institute for Advanced Study in Physics, Zhejiang University, Hangzhou 310027, China
[4]ShanghaiTech Laboratory for Topological Physics, ShanghaiTech University, Shanghai 201210, China
[5]ShanghaiTech Material and Device Lab, ShanghaiTech University, Shanghai 201210, China
[6]Center for High Pressure Science and Technology Advanced Research, Shanghai 201203, China
[7]Beijing National Laboratory for Condensed Matter Physics, Institute of Physics, Chinese Academy of Sciences, Beijing 100190, China
[8]Department of Physics, Clarendon Laboratory, University of Oxford, Parks Road, Oxford OX1 3PU, UK
[9]Key Laboratory for Quantum Materials of Zhejiang Province, Department of Physics, School of Science and Research Center for Industries of the Future, Westlake University, Hangzhou 310030, China
[10]Shanghai Key Laboratory of High-resolution Electron Microscopy, ShanghaiTech University, Shanghai 201210, China
[†]These authors contribute equally to this work.
[*]Correspondence should be addressed to Y.P.Q. (qiyp@shanghaitech.edu.cn) or W.Y.H. (hewy@shanghaitech.edu.cn) or X.L. (linxiao@westlake.edu.cn)

*Contact author: qiyp@shanghaitech.edu.cn, hewy@shanghaitech.edu.cn, linxiao@westlake.edu.cn

**ABSTRACT**. The superconducting diode effect (SDE) is a fascinating nonreciprocal phenomenon where the critical current is different for opposite current directions. It is widely believed that realizing SDE requires breaking both inversion symmetry (IS) and time-reversal symmetry (TRS), which are usually achieved *via* heterostructure engineering and applying external magnetic fields. Here, we report a pressure-induced magnetic-field-free SDE in $NbSe_2$ flakes without any heterostructures. We show that pressure alone breaks the IS, as confirmed by the second harmonic generation. Crucially, upon applying an out-of-plane magnetic field ($B$), the SDE exhibits even-in-B behavior, implying the absence of explicit TRS breaking. This finding challenges the prevailing theoretical paradigm and demonstrates that a magnetic-field-free SDE can emerge without explicitly breaking TRS. Thereby, our work establishes pressure engineering as a powerful tool for inducing nonreciprocal superconductivity and designing versatile, magnetic-field-free superconducting devices.

## I. INTRODUCTION.

The pursuit of non-reciprocal superconducting transport, exemplified by the superconducting diode effect (SDE), has garnered considerable interest in the community of condensed matter and superconducting device physics. Analogous to a semiconductor diode, the SDE enables dissipationless supercurrent flow in one direction while imposing finite resistance in the opposite direction when the critical current is exceeded, offering potential applications in high-efficiency superconducting circuits [1-4] and quantum information technologies [5,6]. Beyond its application promise in superconducting electronics, the SDE also serves as a sensitive probe of fundamental superconducting physics, intertwining with finite momentum pairing [7,8], spin orbit coupling effects [9,10], vortex dynamics [11], and unconventional pairings [12-14]. Owing to its dual connection to both the functional superconducting device and fundamental physics in superconductivity, the SDE has spurred intensive theoretical and experimental investigations [7-35] in recent years.

The first experimental observation of the SDE was reported in the superlattice of Nb/V/Ta under an external magnetic field [9]. Subsequent to this discovery, a substantial number of SDEs were observed by applying an external magnetic field to superconducting systems with no inversion center [10,11,18,21,24,26,27,29,30]. As the SDEs in these noncentrosymmetric systems all exhibit odd symmetry with respect to the applied external magnetic field, these observations are consistent with the prevailing theoretical paradigm [8,15-17] that the SDE requires the simultaneous breaking of both inversion symmetry (IS) and time reversal symmetry (TRS). Recent advances, however, have revealed magnetic field-free SDEs in two distinct categories. The first category observed in magnetism/superconductivity heterostructures [20,36], twisted Moire superconductors [13,34], chiral molecule intercalated $TaS_2$ [14] and unconventional chiral superconducting $CsV_3Sb_5$ [12], align with the conventional requirement of TRS breaking for the SDE, where the TRS is either spontaneously broken or get broken by the proximitized ferromagnets. Strikingly, the second category was reported in van der Waals Josephson junctions [23,32], strained $PbTaSe_2$ [19], high-$T_c$ cuprates [22] and layered FeSe [37] with no direct evidence of TRS breaking. These magnetic-field free SDEs, identified in systems lacking direct experimental signatures of TRS breaking, challenge the long-standing prerequisite of TRS breaking for the SDE [25,28,31,33], prompting a re-examination of the fundamental mechanisms of the SDE.

While prior studies of the SDEs have predominantly focused on intrinsic non-centrosymmetric systems or heterostructures with TRS breaking, here, we report a pressure-induced magnetic field-free SDE in a single-phase $NbSe_2$ flake (Fig. 1B). $NbSe_2$ is originally a centrosymmetric transition metal dichalcogenides (TMDs) where applying pressure allows IS breaking but preserves TRS. In our experiments, the IS breaking in the $NbSe_2$ flake under pressure is confirmed by the second-harmonic generation (SHG), while the magnetic field-even dependence of the SDE gives the hint that the TRS might be preserved. These experimental observations demonstrate that applying pressure can coax a centrosymmetric TMD material into exhibiting nonreciprocal superconductivity, providing a new example of the SDE that occurs in systems without explicitly breaking TRS. Building on these experimental results, we further propose a phenomenological model where the SDE in materials with broken IS originates from the cubic ($\boldsymbol{q}^3$) Cooper pair momentum dependent coupling in the Landau Ginzburg free energy functional. This discovery of a pressure-induced magnetic field-free SDE in $NbSe_2$ flakes redefines the symmetry requirements for the SDE, enriches the known superconducting physics in $NbSe_2$ [38-43], and advances our understanding of superconductivity under symmetry breaking. Furthermore, our high-pressure technique offers a novel strategy to design tunable, magnetic field-free SDEs in quantum materials, advancing the development of programmable superconducting devices.

*Contact author: qiyp@shanghaitech.edu.cn, hewy@shanghaitech.edu.cn, linxiao@westlake.edu.cn

## II. RESULTS

$2H$-$NbSe_2$ is a member of the $2H$ transition metal dichalcogenides that are characterized by a six-fold rotational symmetry in the basal plane and a global IS in the bulk. As illustrated in Fig. 1A, the $2H$ structure comprises two $1H$-$NbSe_2$ monolayers stacked with a 180° rotation. This interlayer stacking aligns the Nb atoms in the upper and lower layers with atomic precision, restoring IS globally in the bulk while breaking IS within individual monolayers. Each layer adopts a distinct in-plane crystallographic direction, armchair in one and zigzag in the other, oriented 30° relative to each other due to the hexagonal lattice geometry [44,45].

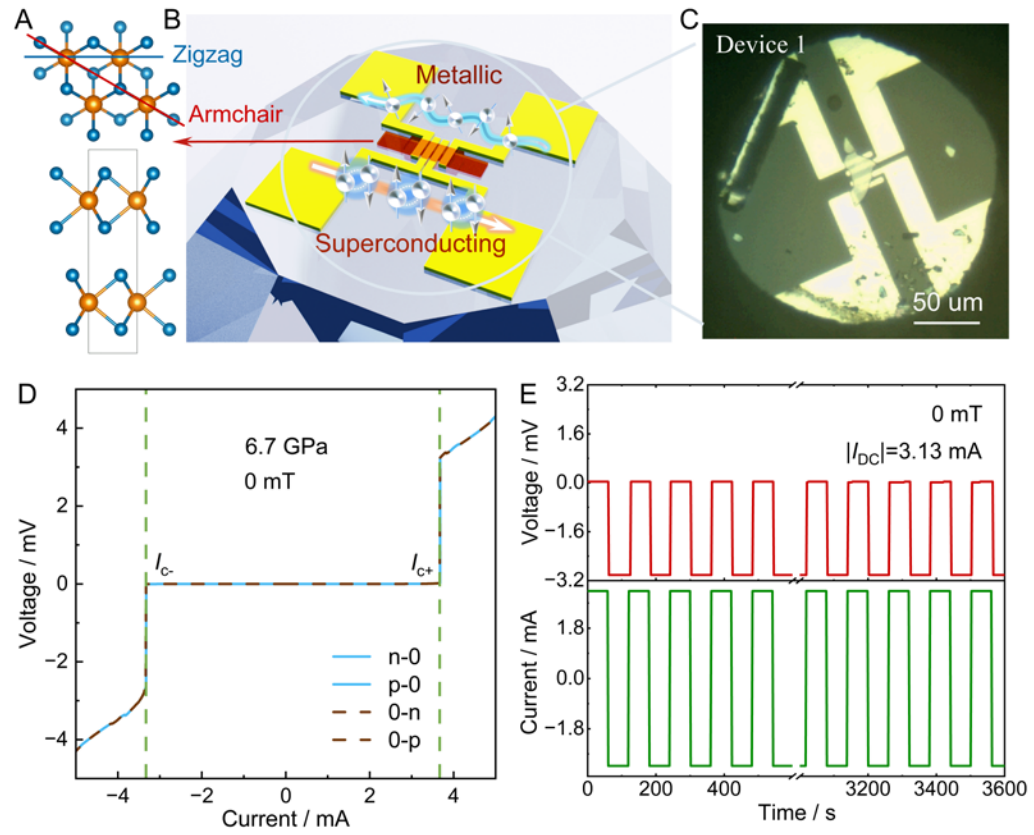

FIG 1. Crystal structure, device configuration and SDE of $NbSe_2$ in high-pressure Device 1. (A) The crystal structures of $2H$-$NbSe_2$ from top and side views. Nb atoms are shown in orange, and Se atoms in blue. (B) The schematic image of the pressure induced SDE on diamond culet. With a current applied, the Cooper pairs flow in one direction while undergoing dissociation in the reverse direction. (C) Optical image of sample chamber in Device 1. (D) *V-I* curve for Device 1 under 6.7 GPa. (E) Half-wave rectification measured at 1.8 K and 11.0 GPa under zero magnetic field. The amplitude of the excitation current is 3.13 mA and the frequency are 0.008 Hz.

Owing to its weak van der Waals interaction between the layers, we mechanically exfoliated micro-sized $NbSe_2$ flakes onto the oxygen-plasma-treated diamond anvil culet with pre-patterned electrodes. The standard four-electrode for the $NbSe_2$ flake devices were fabricated within a diamond anvil cell (DAC) using laser-beam lithography technique. The external circuit was connected outside the DAC *via* Pt foil and Ti/Au (5 nm/35 nm) electrodes (see more details in Methods). Fig. 1C shows an optical image of a typical high-pressure $NbSe_2$ flake Device 1. To determine sample thickness within the DAC, atomic force microscopy was first used to measure samples of varying thickness on $SiO_2$/Si substrate, and then the thickness of DAC-encapsulated samples was estimated *via* the optical contrast under a microscope. The thickness of Device 1 is approximately 10 nm. The electrical transport properties of $NbSe_2$ flakes under pressure are not exactly the same as that of $NbSe_2$ bulk crystals (see more details in Fig. S1 based on analysis from the literature in the Supplemental Material [46] which includes Refs. [47-51]).

The voltage-current (*V-I*) characteristics of Device 1 were measured under high pressures in the absence of an applied magnetic field. To eliminate the effects of residual magnetic flux, the cryostat was warmed above the $T_c$ of the superconducting magnet prior to measurements. The *V-I* curves were acquired *via* a four-step protocol: (1) DC current sweep from zero to positive (0-p); (2) from positive to zero (p-0); (3) from zero to negative (0-n); (4) from negative to zero (n-0). Fig. 1D shows the *V-I* curves at 1.8 K under 6.7 GPa, where the critical currents in the positive and negative directions are marked with dashed lines, revealing a pronounced non-reciprocity ($|I_{c+}| \neq |I_{c-}|$). The superconducting diode efficiency, defined as $\eta = (I_{c+} - |I_{c-}|)/(I_{c+} + |I_{c-}|)$, reaches 4.9%. Notably, the critical current trace (0-p,0-n) coincides with their return trace (p-0, n-0), indicating that the impact of thermal and capacitance effects [31,33] is negligible during the test. The hallmark rectification capability of the superconducting diode effect in Device 1 is demonstrated in Fig. 1E, where a 0.008 Hz square-wave current drives repeated switches between superconducting and resistive normal states. Device 1 exhibited stable rectification over 100 cycles (1 hour), confirming the robustness of the SDE device under high pressure, which is crucial for future applications in superconducting electronics.

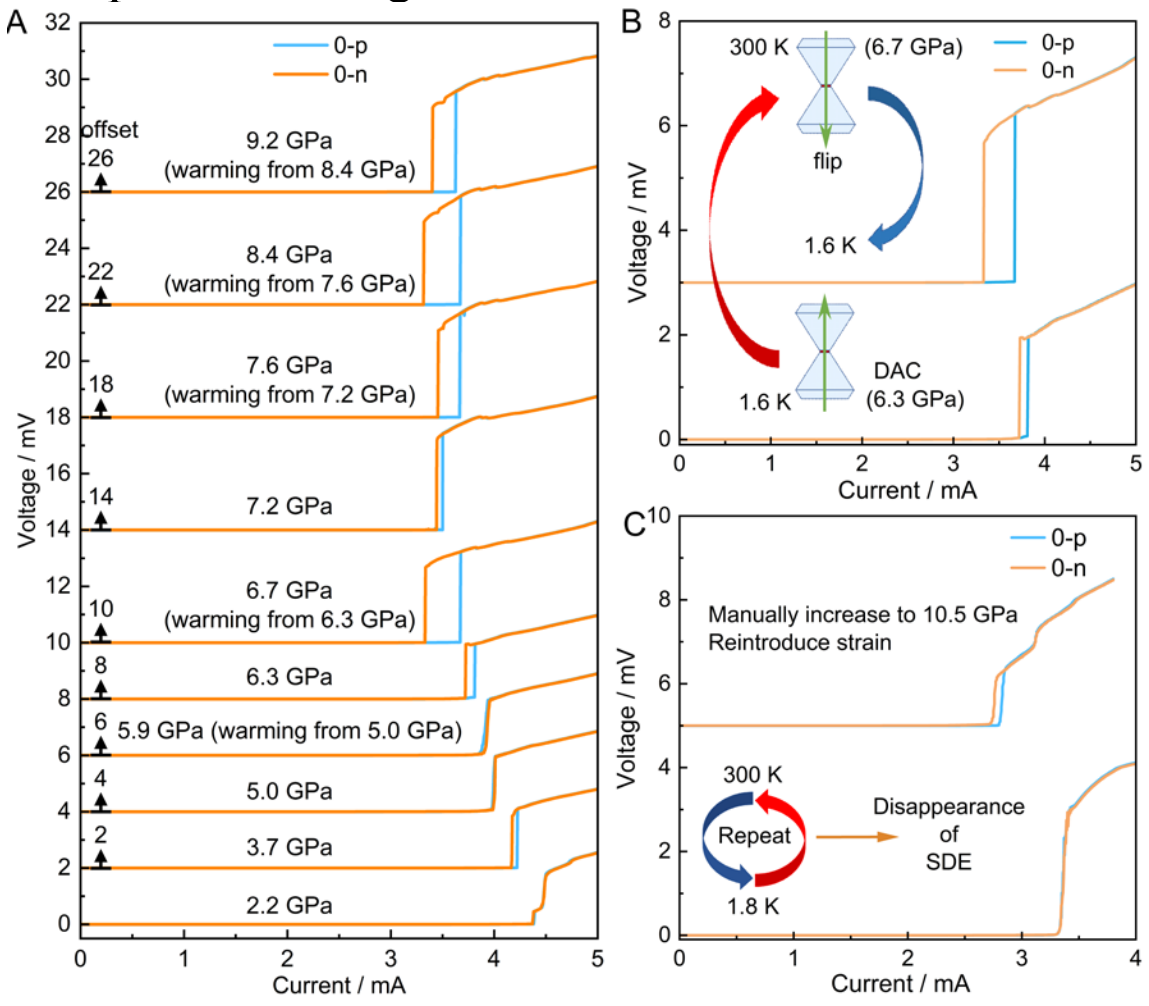

*Contact author: qiyp@shanghaitech.edu.cn, hewy@shanghaitech.edu.cn, linxiao@westlake.edu.cn

FIG 2. Pressure-induced SDE at zero magnetic field of $NbSe_2$ flake in Device 1. (A) *V-I* curves measured by ramping the current from zero to positive (0-p, blue line) and from zero to negative (0-n, orange line) at 1.8 K under various pressures. Parentheses refer to the pressure increase from the cooling-warming cycle. The *V-I* curves are vertically offset for clarity. (B) Comparison of *V-I* curves in Device 1 at 6.3 GPa after flipping 180° in the cryostat. (C) Lower panel: the disappearance of SDE under stable 9.5 GPa and 1.8 K after 6 times of cooling-warming cycles from 7.2 GPa; upper panel: re-emergence of SDE after manually loading pressure to 10.5 GPa at 1.8 K.

To investigate the pressure-dependent evolution of the SDE, Figs. 2A and S2 presents the *V-I* curves of Device 1 as pressure increases from 2.2 GPa to 9.2 GPa. At the initial pressure of 2.2 GPa, the 0-p (positive current sweep) and 0-n (negative current sweep) *V*-I traces overlap, but exhibit a two-step voltage climb, likely due to the incomplete electric contact between the sample and electrodes at low pressure. As pressure increases, mechanical stabilization of the sample electrode interface eliminates this incomplete contact, restoring the conventional single-step of resistance drop in the superconducting transition. Concurrently, a pronounced non-reciprocal critical current emerges ($|I_{c+}|\neq|I_{c-}|$), demonstrating the onset of magnetic field-free SDE at higher pressures.

Upon pressurization to 6.3 GPa, the *V-I* curves for the 0-p and 0-n sweeps start to diverge, with $I_{c+}>|I_{c-}|$, establishing a forward diode polarity in Device 1. Subsequent to this pressure point, after warming to room temperature from 1.8 to 300 K, the calibrated pressure increased to 6.7 GPa compared to that value of 6.3 GPa before putting into the cryostat (See more discussion in Fig. S3). All curves noted with parentheses in Fig. 2A were not subjected to manual pressurization, but only the cooling-warming process. Following the onset of the pressure-induced SDE, the diode efficiency $\eta$ and the asymmetric current $\Delta I_c$ exhibits no monotonic dependence on pressure, but the forward diode polarity remains robustly preserved across all pressures, which suggests that some kinds of implicit mechanism lock the direction of SDE. However, the origin of the fluctuating $\eta$ remains unclear and may stem from inhomogeneous stress distribution within the DAC (as discussed in SI), which is notoriously challenging to quantify. Fig. 2B shows the diode polarity remained unchanged as DAC flipped 180° in cryostat, ruling out residual magnetic flux in the cryostat as the origin of the SDE (detailed discussion in Fig. S3).

Intriguingly, following the final manual pressurization to 7.2 GPa, the pressure within the DAC stabilized at 9.5 GPa after experiencing six cooling-warming cycles, with no further pressure increase observed during subsequent thermal cycling (Fig. S3). Strikingly, the SDE vanished entirely at this stabilized pressure (Fig. 2C), and no SDE could be reinstated even under an external magnetic field (Fig. S4). This behavior starkly contrast with the previously reported SDE in $NbSe_2$ thin film under ambient pressure, where magnetochiral anisotropy under external magnetic fields was considered to drive the effect [10]. This disappearance of SDE at 9.5 GPa highlights a fundamental distinction between the underlying mechanisms of applying pressure and magnetochiral anisotropy. Considering pressure increase may come from the stress release of thermal cycles, the disappearance of SDE could also stem from the uneven-distributed pressure of total stress release after certain thermal cycles. Remarkably, when the stabilized pressure was manually increased to 10.5 GPa, the SDE re-emerged (Fig. 2C), suggesting that the restoration of SDE could be correlated with the re-introduction of stress.

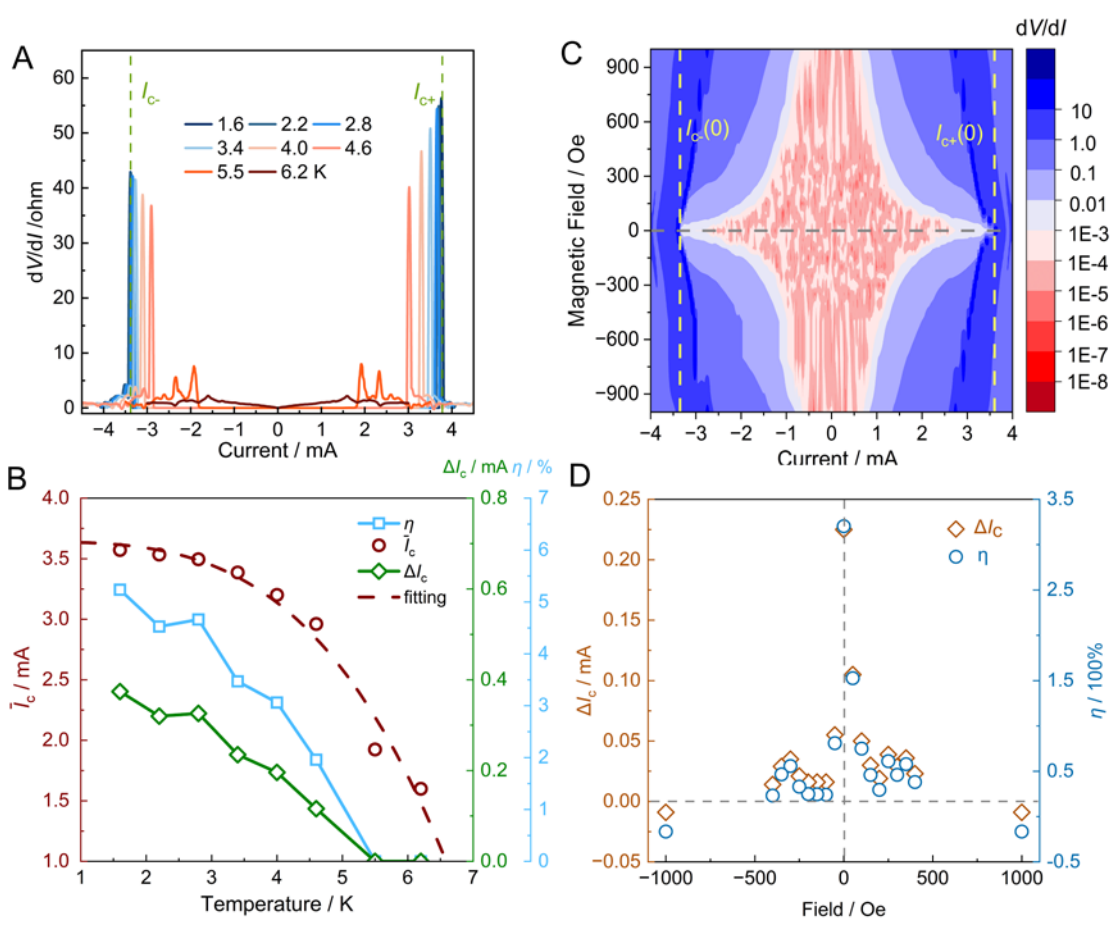

FIG 3. Temperature and magnetic dependence of SDE in Device 1. (A) Differential resistance d*V*/d*I*-*I* curves varying with temperature at 7.6 GPa. The critical current $I_c$ is defined to be the value where the differential resistance reaches its peak in d*V*/d*I*. (B) Temperature dependence of the critical current $I_c$, the asymmetric critical current $\Delta I_c$ ($\Delta I_c=I_{c+}-|I_{c-}|$) and SDE efficiency $\eta$ at 7.6 GPa. (C) Magnetic field dependence of differential resistance d*V*/d*I*-*I* curves at 9.2 GPa. (D) Magnetic field dependence of the asymmetric critical current $\Delta I_c$ and SDE efficiency $\eta$ at 9.2 GPa and 1.8 K.

We further investigated the temperature dependence of the SDE. Fig. 3A and B show the differential resistance (d*V*/d*I*) and critical current ($I_c$) as functions of temperature at 7.6 GPa respectively (see Fig. S5 for analogous data at 7.2 GPa). In the SDE regime, both $I_c$ and the d$V$/d$I$ peaks exhibit pronounced asymmetry, with values significantly enhanced on the positive current side. As temperature increases, the peak broadens progressively and decreases in amplitude, while $I_c$ follows the conventional BCS temperature dependence (Fig. 3B) [52]. Since thermal fluctuations destroy the pairing ground state, $I_c$, $\Delta I_c$, and $\eta$ all decrease to zero as temperature approaches to $T_c$.

*Contact author: qiyp@shanghaitech.edu.cn, hewy@shanghaitech.edu.cn, linxiao@westlake.edu.cn

To investigate the interplay between the SDE and the magnetic field, we measured $V$-$I$ curves at 1.8 K and 9.2 GPa under perpendicular magnetic fields. Both the $I_{c+}$ and $|I_{c-}|$ (defined by the position of differential resistance peak) decrease monotonically with increasing the applied magnetic field (Fig. 3C). Strikingly, however, the forward diode polarity ($I_{c+}>|I_{c-}|$) remains robust across all fields, signifying the magnetic field-free SDE. The asymmetric critical current ($\Delta I_c$) peaks at zero field but diminishes rapidly with applied magnetic fields, exhibiting a symmetric field dependence (Fig. 3D) which contradicts the field-induced SDE in $NbSe_2$ film [10]. This rapid suppression of SDE by magnetic fields can result from the orbital de-pairing effect and a magnetic modulation of the kinetic energy asymmetry between Cooper pairs moving in the opposite directions. The SDE vanishes entirely above 1000 Oe, which is far below the upper critical field, with $\Delta I_c$ reversing sign at both ±1000 Oe (Fig. 3D) which may ascribe to the formation of finite cooper pair momentum [15] (see more details in Discussion). This magnetic field-tunable SDE distinguishes its mechanism from the general odd-in-$B$ behavior, though its origin warrants further investigation.

In Devices 2-5, we observed variations in the onset pressure and magnitude of the SDE, potentially correlated with differences in sample thickness. To rule out artifacts from the contact asymmetry between the sample and the measurement system, we inverted the electrical connection to Device 5 (swapping current/voltage leads at the sample edges). Strikingly, the SDE polarity reversed upon circuit inversion (Fig. S10), confirming that the SDE originates from the intrinsic properties of the $NbSe_2$ flake, instead of external measurement asymmetry. Repeated experiments across multiple devices (Figs. S6-S10) further validate that the pressure-induced SDE is an intrinsic effect in $NbSe_2$ flakes.

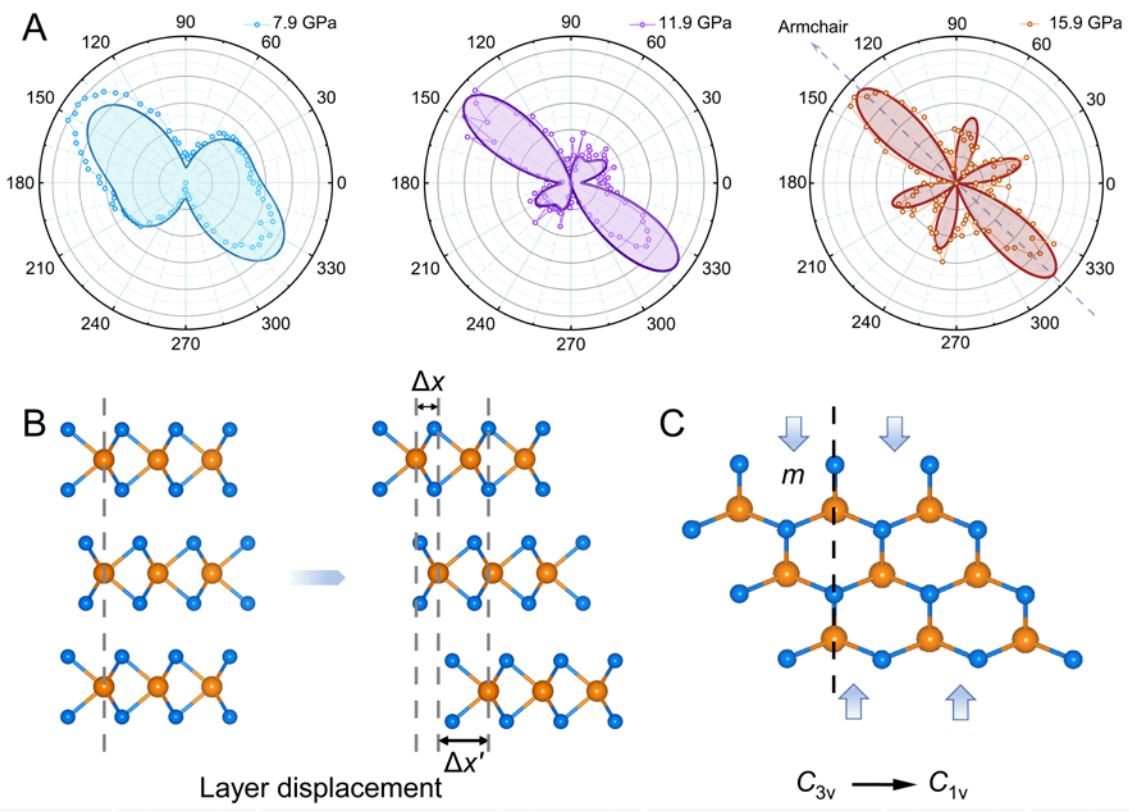

FIG 4. Pressure-induced SHG in $NbSe_2$ flake. (A) Pressure dependence of SHG in 10 nm $NbSe_2$ flake at 7.9 GPa, 11.9 GPa and 15.9 GPa, respectively. (B) Layer displacement leading to the breaking of IS, $\Delta x \neq \Delta x'$. (C) Degradation of $C_{3v}$ to $C_{1v}$ leaving one mirror symmetry along armchair direction.

Generally, the emergence of SDE requires IS breaking. However, $NbSe_2$ has a centrosymmetric structure that preserves IS. This implies that applying pressure to $NbSe_2$ thin films must induce IS breaking. To confirm this, we performed high-pressure second-harmonic generation (SHG) measurements. Fig. 4A and Fig. S11 illustrate the evolution of the SHG intensity as a function of pressure. It is almost negligible under low pressures. However, as pressure increases to 5.7 GPa, a discernible SHG signal emerged. In the initial appearance of SHG, although the dominant signal remains along the armchair direction, there are no clear boundaries between the strongest and the second-strongest signals, showing a butterfly-like two-fold pattern as shown at 7.9 GPa. Only when the pressure reaches 11.9 GPa, a four-fold pattern becomes distinct. The SHG intensity peaks along the crystallographic armchair direction, with a secondary maximum at approximately 60º deviation. It is noteworthy that the critical pressure for the emergence of the SHG and its pattern is highly related to the sample's thickness based on analysis from the literature in the Supplemental Material [51]). Above 20 GPa, the SHG signal was suppressed and vanished.

Combining the phenomenon that SDE will disappear after experiencing multiple thermal cycles, the origin of SHG signal may be attributed to the stress induced by the uneven-distributed pressure inside DAC, which can be tuned through thermal cycling. Crucially, the pressure-induced SHG in $NbSe_2$ flake is not local but can be detected across the surface of uniform samples, indicating that IS breaking is not a random phenomenon in particular regions but occurs globally throughout the sample. These SHG results confirm that IS is broken in the pressurized $NbSe_2$ flakes that exhibit SDE, with plausible mechanisms including inter-layer sliding and in-plane bonding stretching (Fig. 4B and C). The lattice bonding stretching further disrupts the original three-fold rotational symmetry. In the pressure range of 5.7 GPa to 15.9 GPa, the SHG intensities are featured by the predominant peaks (at 135$^{o}$ and 315$^{o}$) along the armchair direction, suggesting a preserved symmetry in the armchair direction. We assume that the mirror symmetry along the armchair direction is approximately retained under the lattice bonding stretching (Fig. 4C), meaning that the lattice symmetry of $NbSe_2$ layers is reduced from the $C_{3v}$ to the $C_{1v}$ point group. Employing the $C_{1v}$ symmetry, we successfully performed the simulation of the SHG pattern of a 10 nm $NbSe_2$ flake under 15.9 GPa (see the detailed fitting in Supplementary information). The simulation fits well with the experimental results (Fig. 4A). Importantly, the dominant SHG signals observed along the armchair direction indicate a net pressure-induced electric polarization, which has been thought to play critical roles in the SDE without explicitly breaking TRS [19]. Notably, the supercurrent direction in our devices (Fig. S12) is found to be almost parallel to the electric polarization direction, strongly supporting an intrinsic correlation between the SDE and electric polarization. However, due to the challenges associated with characterizing flakes in DAC by synchrotron XRD, further studies on the pressure-induced structural distortion are imperative.

*Contact author: qiyp@shanghaitech.edu.cn, hewy@shanghaitech.edu.cn, linxiao@westlake.edu.cn

## III. DISCUSSIONS

The discovery of a magnetic field-free SDE in pressurized 2*H*-$NbSe_2$ flakes represents a pivotal advance in nonreciprocal superconductivity. Conventionally, SDE requires the breaking of TRS *via* applied magnetic fields or proximity to magnetic orders. Strikingly, we demonstrate that applying pressure alone induces SDE in single-phase superconducting $NbSe_2$ flakes without explicit breaking of TRS. This observation challenges the prevailing theoretical frameworks that posit simultaneous IS and TRS breaking as prerequisites for SDE, thereby demanding new understandings for the mechanisms underlying the SDE.

The requirement of TRS breaking in prior theoretical frameworks arises from the helical pairing state, [8,15-17] where the pairing order parameter is expressed as $\Delta(\boldsymbol{r}) = \Delta e^{i\boldsymbol{q}_0\cdot\boldsymbol{r}}$ . Here, the finite Cooper pair momentum $\boldsymbol{q}_0$ in the phase factor breaks TRS. In such states, the Landau-Ginzburg free energy density $f(\Delta_{\boldsymbol{q}})$ is minimized at finite $\boldsymbol{q}_0$. Near this equilibrium pairing momentum $\boldsymbol{q}_0$, IS breaking introduces a cubic term $(\boldsymbol{q}$-$\boldsymbol{q}_0)^3$ into $f(\Delta_{\boldsymbol{q}})$, which breaks the symmetry of $f(\Delta_{\boldsymbol{q}})$ about $\boldsymbol{q}$=$\boldsymbol{q}_0$. The supercurrent density $\mathrm{j}(\boldsymbol{q}) = df(\Delta_{\boldsymbol{q}})/d\boldsymbol{q}$ [8,15-17] inherits this asymmetry due to the $(\boldsymbol{q}$-$\boldsymbol{q}_0)^3$ term, resulting in unequal critical currents for opposing directions. Crucially, this analysis reveals that the $\boldsymbol{q}^3$ terms in the Landau Ginzburg free energy density plays the essential role in generating the SDE.

Building on this insight, we propose a minimal Landau-Ginzburg free energy density $f(\Delta_{\boldsymbol{q}})$ that incorporates a $\boldsymbol{q}^3$ term coupled to IS breaking. Here, IS breaking is manifested by an in-plane electric polarization along the armchair direction as suggested by the SHG simulation (see simulations in the Supplemental Material and literatures [47,48]). To preclude helical finite momentum pairing, the linear $\boldsymbol{q}$ term is explicitly excluded. Besides, there may exist other kinds of finite momentum pairing [43,53] considering that the field could tune the SDE direction in higher magnetic fields (Fig. 3D). In this model, the free energy density $f(\Delta_{\boldsymbol{q}})$ is minimized at $\boldsymbol{q}$=$\boldsymbol{0}$, ensuring that Cooper pair condense at zero momentum in thermal equilibrium. At $\boldsymbol{q}$=0, the $\boldsymbol{q}^3$ term induced asymmetry of $f(\Delta_{\boldsymbol{q}})$ imposes different energy costs for Cooper pairs moving in opposite directions, directly yielding distinct critical supercurrents $I_{c+}$ and $I_{c-}$. This phenomenological Landau Ginzburg model provides a basic mechanism for the magnetic field-free SDE observed in pressurized $NbSe_2$ flakes, though the microscopic origin of the $\boldsymbol{q}^3$ term, potentially tied to pressure-induced IS breaking, requires further investigation. By assuming the critical role of in-plane electric polarization, we expect that this asymmetric $f(\Delta_{\boldsymbol{q}})$ at $\boldsymbol{q}$=0 is generally applicable to quasi-two-dimensional superconductors with an in-plane polar axis ($C_1$, $C_{1v}$, $C_2$ and $C_{2v}$ symmetry [54]). Crucially, symmetry constraints on $f(\Delta_{\boldsymbol{q}})$ require the coupling coefficient of the $\boldsymbol{q}^3$ term to reverse sign under time reversal. This requirement implies that in addition to IS breaking, the $\boldsymbol{q}^3$ term in $f(\Delta_{\boldsymbol{q}})$ must arise from either an implicit TRS breaking within the system (implicit TRS breaking means that the order breaks TRS has no direct linear coupling to external out-of-plane magnetic fields) [19,23] or a nonequilibrium dynamical coupling to the externally driven current itself [55-57]. Considering that monolayer $NbSe_2$ with $C_3$ symmetry is close to a ferromagnetic instability [58], it is possible that after applying pressure, an in-plane magnetic order spontaneously arises from our

*Contact author: qiyp@shanghaitech.edu.cn, hewy@shanghaitech.edu.cn, linxiao@westlake.edu.cn

$NbSe_2$ flakes. Such in-plane magnetic order, which may be either ferromagnetic or anti-ferromagnetic [58,59], is coupled quadratically to an external out-of-plane magnetic field and therefore exhibits the implicit TRS breaking. To pin down the specific magnetic order underlying the observed SDE, further transport measurements to detect the hysteretic magneto-resistance under an in-plane magnetic field would be instrumental.

## IV. CONCLUSIONS

Our discovery of pressure-induced SDE with even-in-B behavior challenges the prevailing theoretical paradigms and necessitates a revised theoretical framework to account for the SDE without explicitly breaking time reversal symmetry. Future studies integrating experimental works with advanced theoretical modeling, could unravel the interplay of lattice distortions, electronic correlations and Cooper pair dynamics underlying the pressure induced magnetic-field-free SDE. By bridging emergent symmetry-breaking physics with pressure engineered quantum materials, our work establishes a blueprint for designing field-free nonreciprocal devices, heralding a new era of scalable, energy-efficient superconducting electronics and quantum technologies.

## ACKNOWLEDGMENTS

This work was supported by the National Key R&D Program of China (Grant No. 2023YFA1607400) and the National Natural Science Foundation of China (Grant Nos. 52272265, 12474018, 12404161). W.Y.H. acknowledges the support from the National Natural Science Foundation of China (Grant No. 12304200), the BHYJRC Program from the Ministry of Education of China (Grant No. SPST-RC-10), the Shanghai Rising-Star Program (24QA2705400), the Science and Technology Commission of Shanghai Municipality (No. 24LZ1401000) and the start-up funding from ShanghaiTech University. X.L. acknowledges the support from the National Key Research and Development Program of China (Grant No. 2024YFA1408101), National Natural Science Foundation of China (Grant No. 12474131), “Pioneer” and “Leading Goose” R&D Program of Zhejiang (Grant No. 2024SDXHDX0007) and Zhejiang Provincial Natural Science Foundation of China for Distinguished Young Scholars (Grant No. LR23A040001). T.L. acknowledges the support from the National Natural Science Foundation of China (Grant No. 92565201) and the Fundamental Research Funds for the Central Universities. The authors thank the Analytical Instrumentation Center (# SPST-AIC10112914), SPST, ShanghaiTech University. This work was carried out with the support of ShanghaiTech Material and Device Lab (SMDL20191219), ShanghaiTech University.

*Data availability*—There are no publicly available research data or software supporting this manuscript. Requests for further information or data should be sent to the authors.

*Contact author: qiyp@shanghaitech.edu.cn, hewy@shanghaitech.edu.cn, linxiao@westlake.edu.cn

*Contact author: qiyp@shanghaitech.edu.cn, hewy@shanghaitech.edu.cn, linxiao@westlake.edu.cn

*Contact author: qiyp@shanghaitech.edu.cn, hewy@shanghaitech.edu.cn, linxiao@westlake.edu.cn

*Contact author: qiyp@shanghaitech.edu.cn, hewy@shanghaitech.edu.cn, linxiao@westlake.edu.cn